# Study of strain effect on in-plane polarization in epitaxial BiFeO$_3$ thin films using planar electrodes

Zuhuang Chen,[1,*] Xi Zou,[1] Wei Ren,[2] Lu You,[1] Chuanwei Huang,[1] Yurong Yang,[2,3] Ping Yang,[4] Junling Wang,[1] Thirumany Sritharan,[1] L. Bellaiche,[2] and Lang Chen[1,†]

[1] School of Materials Science and Engineering, Nanyang Technological University, Singapore, 639798, Singapore

[2] Institute for Nanoscience and Engineering and Physics Department, University of Arkansas, Fayetteville, Arkansas 72701, USA

[3] Physics Department, Nanjing University of Aeronautics and Astronautics, Nanjing 210016, China

[4] Singapore Synchrotron Light Source (SSLS), National University of Singapore 5 Research Link, 117603, Singapore

* glory8508@gmail.com

† langchen@ntu.edu.sg

**Abstract**

Epitaxial strain plays an important role in determining physical properties of perovskite ferroelectric oxide thin films. However, it is very challenging to directly measure properties such as polarization in ultrathin strained films using traditional sandwich capacitor devices, because of high leakage current. We employed a *planar electrode* device with different crystallographical orientations between electrodes along *different* electric field orientation to directly measure the in-plane polarization-electric field (P-E) hysteresis loops in fully strained thin films. At high misfit strains such as -4.4%, the pure Tetrogonal-like phase is obtained and its polarization vector is constrained to lie in the (010) plane with a significantly *large* in-plane component, *~44 µC/cm$^2$*. First-principle calculations are carried out in parallel, and provide a good agreement with the experimental results. Our results pave the way to design in-plane devices based on T-like BFO and the strategy proposed here can be expanded to study all other similar strained multiferroic ultrathin films.

The effect of substrate-induced, epitaxial strain on ferroelectric properties of thin films has received a keen research interest in recent years, not only for fundamental reasons, but also for practical applications.[1] Published research has shown that the polarization,[2] dielectric constant,[3] piezoelectricity,[4] domain structure[5,6] and phase stability[4,7] of these films can be modified by strain because of the inherent coupling between the elastic and ferroelectric phenomena. However, direct experimental study in fully-strained thin films on the effect of strain on ferroelectric properties, such as polarization, is challenging. This is because the critical thickness for the appearance of strain-relieving dislocations is normally below a few tens of nanometers and thus leakage current in the sandwich capacitors is usually high.[1] Furthermore, it is found that the presence of dislocations alters the physical properties of the film, making its ferroelectric properties inhomogeneous and often degraded.[8,9] Therefore, most reported works are restricted to low-strain systems where the lattice mismatch with the underlying substrate is small (< 1%) in order to increase the critical thickness and thus minimize leakage currents. This has impeded a more comprehensive study of the intrinsic effect of strain on the polarization.[2,10] Furthermore, it is known that the determination of in-plane polarization and exact polarization direction of ferroelectric ultrathin films under different strain states is important not only from the point of view of fundamental physics, but also from viewpoint of practical applications, such as ferroeletric photovoltaic device.[11,12] However, there are no systematic experiments on the determination of in-plane polarization components in epitaxial ferroelectric thin films under different strain states. This fact hinders a comprehensive understanding of the overall polarization vectors and projections on different axis in these complex ferroelectric oxides. In this paper, we proposed a general strategy to study the strain-dependent polarizations, particularly the in-plane components. We used BiFeO$_3$ (BFO) as a model system to better understand the direction and magnitude of the in-plane polarization component. Interestingly, this method can be extended to study other rhombohedral/monoclinic ferroelectric materials with equilibrium domain structure, such as $PbZr_{0.5}Ti_{0.5}O_3$ thin films.

Among ferroelectrics, BFO is of particular interest due to its lead-free nature, large polarization, room-temperature multiferroism and domain wall functionalities.[13,14] In 2003, an unexpectedly large polarization (~ 60 μC/cm$^2$) was recorded in (001) epitaxial BFO film grown on $SrTiO_3$ (STO) substrate, and was considered to be due to the strain effect.[15] (note that pseudocubic notations are used in this paper) However, a later study revealed that polarization in strain-free, high-quality, rhombohedral BFO single crystals is similar to that in the epitaxial film,[16] and the polarization in BFO was then treated as strain-insensitive.[16, 17] This claim was partly supported by independent theoretical calculations, which confirmed that the spontaneous polarization in the R-like BFO phase is intrinsically large.[18,19,20] Subsequent studies further indicated that the epitaxial strain induces a rotation of the polarization vector rather than enhancing its magnitude.[18, 21] First-principle calculations, on the other hand, predicted a new, metastable tetragonal phase of BFO having a large c/a ratio of ~1.27, whose spontaneous polarization was estimated to have the giant value of ~150 μC/cm$^2$.[18] Search for this new phase was met with success recently. Indeed, studies on epitaxial BFO films grown on substrates with lattice mismatches exceeding ~-4%, such as $LaAlO_3$ (LAO), have detected the existence of a tetragonal-like (T-like) phase at such large strains.[4, 7] Subsequent detailed crystal structure analyses confirmed that this T-like phase is not exactly tetragonal, but is a monoclinic $M_C$ phase, whose polarization vector is along a [$u0v$] direction in the (010) plane.[6, 22] It was also shown that BFO film with pure T-like phase could be obtained on LAO substrate, only for film thicknesses less than ~30 nm.[4, 23, 24] For larger film thicknesses, strain relaxation occurs by structural changes leading to the coexistence of multiple phases in those films.[4, 23] Therefore, direct polarization-electric field (*P-E*) hysteresis loop measurement for the pure T-like phase has not been forthcoming because of the high leakage currents in such ultra-thin films when one uses the conventional sandwich capacitor device. Very recently, the out-of-plane polarization of the T-like phase was measured on a series of strain-relaxed mixed-phase films grown on LAO, and was deduced to be ~150 $\mu C/cm^2$.[25] Such a giant out-of-plane polarization was further supported by a subsequent experiment using high resolution scanning transmission electron microscopy, based on the relatively large ionic displacements.[26] However, the *in-plane* component of the polarization in a BFO film with pure T-like phase has not been experimentally measured to the best of our knowledge. As a result, the full characterization of the spontaneous polarization vector about its direction and magnitude, in the T-like phase is not experimentally established yet.

~30-nm-thick BFO films were grown by pulsed laser deposition on (001)-oriented STO, $(LaAlO_3)_{0.3}(Sr_2AlTaO_6)_{0.7}$ (LSAT) and LAO substrates which provide lattice mismatches of -1.4%, -2.4% and -4.4%, respectively.[22] The deposition temperature and the oxygen pressure were 700 °C and 100 mTorr, respectively.[27] Following the BFO deposition, 40 nm Pt planar electrodes were patterned on top via a standard lift-off procedure. $\theta - 2\theta$ X-ray diffraction (XRD) patterns were obtained using a Rigaku four-circle x-ray diffractometer. XRD reciprocal space mapping (RSM) was done at Singapore Synchrotron Light Source ($\lambda$ = 1.5405 Å). The thickness of the BFO films was calibrated by synchrotron X-ray reflectivity, as shown in the supplementary material.[28] The topography and piezoelectric force microscopy (PFM) were carried out on an Asylum Research MFP-3D atomic force microscope (AFM). *P-E* loops were measured at room temperature, at a frequency of 1 kHz using a Precision LC ferroelectric tester (Radiant Technologies). Reproducibility was confirmed by measuring several *P-E* loops in at least five different capacitors.

Planar electrode device has been used to investigate the in-plane domain switching in BFO films.[29,30,31] In order to determine the polarization direction in the pure T-like phase, and also to shed more light on the effect of strain on polarization, the planar Pt electrodes were patterned on the strained BFO films to directly measure the in-plane *P-E* loops using remnant hysteresis measurement.[28] As shown schematically in **Figure. 1**(a), the edge of the Pt electrode was aligned along the <100> and <110> directions of the substrate enabling the application of an electric field in different crystallographic directions. According to finite element analysis, majority of the film within the gap between the two electrodes experiences almost constant in-plane electric field.[28, 31] The nominal applied electric field $E$ was determined using $E=V/d$, where $V$ is the applied voltage and $d$ is the channel width. The nominal polarization $P$ was calculated using $P=Q/hl$, where $Q$ is the polarization charge, $h$ is the BFO film thickness and $l$ the length of the electrode.

We carried out first-principles calculations of BFO films under different epitaxial strains using the Vienna *ab-initio* simulation package (VASP)[32] to better understand the strain-polarization coupling. Density-functional calculations within the local spin density approximation (LSDA) plus the Hubbard parameter U (with U=3.8 eV for Fe ions) are performed. Fe ions that are the nearest neighbors were imposed with opposite magnetic moments –consistent with the G-type antiferromagnetic order known to exist in BFO. We used the projector augmented wave (PAW) method[33] with an energy cutoff of 550 eV and a 3×3×3 k-point mesh. Three different directions were examined for the polarization: along [111] (*R3c* space group), [*uuv*] directions ($v > u$, corresponds to monoclinic $M_A$) and [*u0v*] directions (corresponds to monoclinic $M_C$). 40-atom supercells were used to study all these structures, and the computations included oxygen octahedral tilting. The total polarization of each simulated structure was calculated from the modern theory of polarization based on Berry phase.[34]

**Figure 1(b)** shows the results of X-ray diffraction (XRD) *θ*-2*θ* measurements. Only the (00*l*) peaks from the substrate and the film are evident, indicating that all films are epitaxially grown on the substrates. It can be inferred from the out-of-plane *c* lattice parameters that the films on the low-mismatch substrates of STO and LSAT have R-like structures while that on high-mismatch LAO has a pure T-like structure.[22] The thickness fringes around the Bragg peaks indicate that all films have a high crystallinity and a smooth surface, which was also confirmed by AFM topography imaging, where atomically flat terraces were a clear feature in AFM topography images as evidenced in the example shown in **Figure 1(c)** for the film grown on LAO. Previous studies have reported that BFO films grown on STO and LSAT are coherently strained when the film thickness is less than ~40nm.[35,36,37] To probe the strain state, XRD reciprocal space mapping (RSM) measurements were carried out. Typical RSM of the ($\bar{1}$03) reflection for the film grown on LSAT is shown in **Fig. 1(d)**. The peaks from the substrate and film have identical in-plane positions indicating perfect epitaxy, and that the film is fully strained, as consistent with a previous report.[36] The absence of splitting in ($\bar{1}$03) peak of the BFO suggests that the unit-cell shape of the film appears to be *tetragonal,* which seems at odds with the inferences of the R-like phase deduced from the *c*-lattice values shown in **Figure 1**. **Figure 2**(a) shows the in-plane *P-E* hysteresis loops for the film on LSAT substrate when *E* is along the [100] direction. The loops are saturated and rectangular in shape, with a polarization of $36\pm3$ $\mu$C/cm$^2$ and a nominal coercive field of ~75 kV/cm. This directly proves that the film on LSAT has a finite in-plane polarization component. This was further confirmed by in-plane piezoelectric force microscopy (PFM) images (not shown) where a weak, but detectable piezoelectric signal was obtained. The existence of in-plane polarization components seems *incompatible* with the ($\bar{1}$03) RSM result (**Figure 1d**) which indicates that the film may be tetragonal. Such a decoupling between the external lattice and the internal polar symmetry has been argued previously in BFO films.[13, 35, 37] As pointed by Catalan and Scott,[13] it is possible that in ultrathin films the external lattice could appear to be tetragonal shape, while internal polar symmetry retains to be monoclinic that allows for a polarization with in-plane component to exist.[13] This interesting phenomenon is worth further detailed studies in future works from both theoretical and experimental points of views but is not within the main scope of this paper.

**Figure 2(b)** shows the *P-E* hysteresis loop for the pure T-like film on LAO substrate when *E* is along the [110] direction. Square-shaped saturated hysteresis loops with saturated polarization of $32\pm2$ $\mu C/cm^2$ and coercive field of ~150 kV/cm are obtained. Previous studies have shown that the R-like phase is monoclinic $M_A$ phase with a polarization vector along a [*uuv*] direction in the ($1\bar{1}0$) plane while the T-like phase is monoclinic $M_C$ phase with a polarization vector along a [*u0v*] direction in the (010) plane in relatively thick BFO films using synchrotron X-ray RSM and PFM.[6, 22] In order to determine the exact polarization vector and to confirm the polar symmetry of these strained ultrathin films, we studied the effect of in-plane *E* orientation on the polarization. **Figures 2(c) and 2(d)** show the *P-E* loops of the films grown on STO (R-like) and LAO (T-like) with *E* aligned along <100> and <110> directions, respectively. Hysteresis loop saturation in each direction is evident. The remnant polarization of the film on STO is $45\pm3$ $\mu C/cm^2$ for *E* lying along [100], as consistent with our previous study,[31] and $33\pm2$ $\mu C/cm^2$ for *E* applied along [110]. It is interesting to note that the [100] remnant polarization of the R-like film, $P_{E//[100],R}$ is approximately $\sqrt{2}$ times the $P_{E//[110],R}$. While in the film grown on LAO, $P_{E//[110],T}$ which is $32\pm2$ $\mu C/cm^2$ is approximately $\sqrt{2}$ times $P_{E//[100],T}$ which is $22\pm2$ $\mu C/cm^2$. These observations confirm that the R-like and T-like phases have distinct polar symmetries and their in-plane polarization projections are rotated 45° from each other.

To make sure that in-plane domain switching is the only source of the P-E loops obtained with the planar electrode geometry, we have performed PFM studies to investigate domain structure change before and after applying the in-plane electric field to switch the BFO film within the channel. As a proof, we study in-plane domain switching of BFO films grown on STO substrates. Figure 3(a) schematically shows the unit cell of the R-like film with its possible polarization vectors, and Figure 3b shows the in-plane domain configurations of each stripe domain. As shown in Figures 3(c)-(e), when an electric field larger than the coercive field is applied along a <100> direction to the film, the original intersected stripe domain network is converted to a highly-aligned stripe domain pattern composed of only two domain variants with the in-plane net polarization following the applying field direction. Note that during the switching process by applying in-plane electric field, the out-of-plane PFM images (not shown) remain in the same tone, implying that there are no change in out-of-plane polarization and only in-plane polarization rotations occur. A complete switching of all the in-plane domains is observed when the applied electric field is reversed. These observations indicate that the polarization what we measured in Fig. 2(c) is totally from the in-plane domain switching. For $M_A$-type R-like domains, the in-plane polarization in each individual domain is along <110> and the net in-plane polarization of stripe domains is the sum of their resolved components in <100>. Thus, the measured polarization ($P_{E//[100],R}$) can be seen as the projection of the in-plane component of the spontaneous polarization ($P_{in,R}$) along the chosen <100> direction, that is: $P_{E//[100],R} = \frac{\sqrt{2}}{2} P_{in,R}$. As shown in Figures 3(f)-(h), when $E$ is applied along <110> direction, the in-plane polarizations of half populations of the domains are perpendicular to $E$ while the other half are parallel. Those perpendicular ones will not contribute to the measured polarization. Therefore, only 50% of the in-plane polarizations of the total R-like domains will contribute to the polarization $P_{E//[110],R}$ measured by the in-plane P-E loop. Therefore, the in-plane polarization of each domain in the R-like film is then $P_{in,R} = 2P_{E//[110],R} = \sqrt{2}P_{E//[100],R}$, which is consistent with the above results of in-plane P-E loops measurement of R-like films. Consider the actual in-plane polarization orientations of T-like ($M_C$) and R-like ($M_A$) phases are rotated 45° from each other, the in-plane component of polarization of T-like phase obtained in the film on LAO is determined to be $P_{in,T} = 2P_{E//[100],T} = \sqrt{2}P_{E//[110],T} \sim 44\pm4$ $\mu C/cm^2$, which is nearly *twice* as large as the total polarization of a robust ferroelectric BaTiO$_3$ single crystal.[2] The in-plane polarization of the T-like phase of BFO films is therefore significant in comparison to commercial ferroelectrics. Using the value of 150 $\mu C/cm^2$ reported in the literature for the out-of-plane component of polarization of this phase,[25] we may infer that its polarization vector (which is constrained to the (010) plane) is rotated away by ~16° from the [001] direction.

Figure 4(a) shows the in-plane *P-E* loops of the strained films on different substrates. It is clear that the in-plane polarization value of the R-like phase reduces with increasing in-plane compressive strain. As a matter of fact, this in-plane component is $P_{in,R} = \sqrt{2}P_{E//[100],R}$ ~ $\sqrt{2} * 36$ $\mu C/cm^2$ ~ 50 $\mu C/cm^2$ for BFO films grown on LSAT, while it is $P_{in,R} = 2P_{E//[110],R} = \sqrt{2}P_{E//[100],R}$ ~ $\sqrt{2} * 45 \mu C/cm^2$ ~ 64 $\mu C/cm^2$ for BFO films grown on STO. Such decrease has been reported in various theoretical studies [19, 20, 38] and is due to the strain-induced polarization vector rotation -- that is, the larger compressive strain in the film on LSAT will rotate the polarization vector towards the [001] direction resulting in a smaller in-plane component (see **Figure 4(b)**).[21] We find that the coercive field increases with increasing compressive strain, which leads to a larger c/a ratio. This is consistent with a previous report on strain-dependent coercivity in BFO films on a piezoelectric substrate of (1-*x*)Pb(Mg$_{1/3}$Nb$_{2/3}$)O$_3$–*x*PbTiO$_3$ where such dependence was observed too.[39]

To better understand the fundamental behavior and values of the in-plane components of the polarizations, we also carried out first-principles calculations of BFO under different epitaxial strains. **Table 1** illustrates the computed Cartesian components of the resulting polarization vectors. As we increased the in-plane compressive strain (i.e., decreased the in-plane lattice parameters), the in-plane polarizations (in the *x* and *y* directions) gradually reduced in magnitude (when going from *R*3*c* to $M_A$) while the out-of-plane component (in the *z* direction) increased. This result reveals that the polarization vector constrained to the (1$\bar{1}$0) plane indeed rotates from [111] towards the [001] direction with increasing the compressive strain.[19, 20, 38] When the strain magnitude is large enough (i.e., beyond -4% here), the R-like phase is transformed to the T-like phase resulting in an abrupt increase of the out-of-plane polarization. Nonetheless, we note that the predicted in-plane polarizations are still substantial, even for the T-like phase, which are therefore measurable experimentally. In addition, we compare the energies of both the $M_A$ and $M_C$ phases at the high strains of -4.4 to -4.5% and find that the $M_A$ phase is energetically more favorable than the $M_C$ phase, but by a small energy of only 11.4 meV/formula unit. Hence, it is plausible that polydomain structures may form, or the energy balance could be tipped in favor of the $M_C$ phase by surface effects not incorporated in the present calculation.[24] It is noteworthy to realize that the present calculations give an in-plane polarization component of 43 $\mu$C/cm$^2$ in the $M_C$ phase, which is larger than those in previous theoretic reports based on *Cc*/*Cm* symmetry.[19,20,26,32] Our calculated results imply that the resultant polarization vector subtends an angle of 17.2° with the [001] out-of-plane direction. These predictions are in excellent agreement with the experimentally measured values reported in this paper.

In summary, we developed a planar electrode technique to directly measure the in-plane component of the spontaneous polarization in ultrathin epitaxial strained BFO films which facilitated the characterization of the effect of misfit strain on the polarization of the epitaxial films. Our results reveal that the polarization vector in R-like phase is constrained to lie within (110) plane and rotates from [111] toward [001] direction with increasing compressive strain; and the in-plane component of the spontaneous polarization in the pure T-like phase in the film grown on LAO is as large as $44\pm4$ $\mu$C/cm$^2$, and that the resultant polarization vector, which lies on the (010) plane, is rotated away by around 16° from the [001] direction. The coercive field is found to increase with increasing compressive strain. The experimental results were substantiated by first-principles modeling. The in-plane device structure can also be used to investigate the in-plane polarization and strain effect of other ferroelectric and multiferroic films with in-plane components.

**Acknowledgements** ZC acknowledges the support from Ian Ferguson Postgraduate Fellowship. LC acknowledges the supports from the Singapore National Research Foundation under CREATE program: Nanomaterials for Energy and Water Management. The computational work of W.R., Y.Y. and L.B. is financially supported by the Department of Energy, Office of Basic Energy Sciences, under contract ER-46612. These authors also thank ARO grant W911NF-12-1-0085, NSF grants DMR-1066158 and DMR-0701558 and ONR Grants N00014-11-1-0384 and N00014-08-1-0915 for discussions with scientists sponsored by these grants. Some computations were also made possible thanks to the MRI grant 0722625 from NSF, the ONR grant N00014-07-1-0825 (DURIP) and a Challenge grant from the Department of Defense.

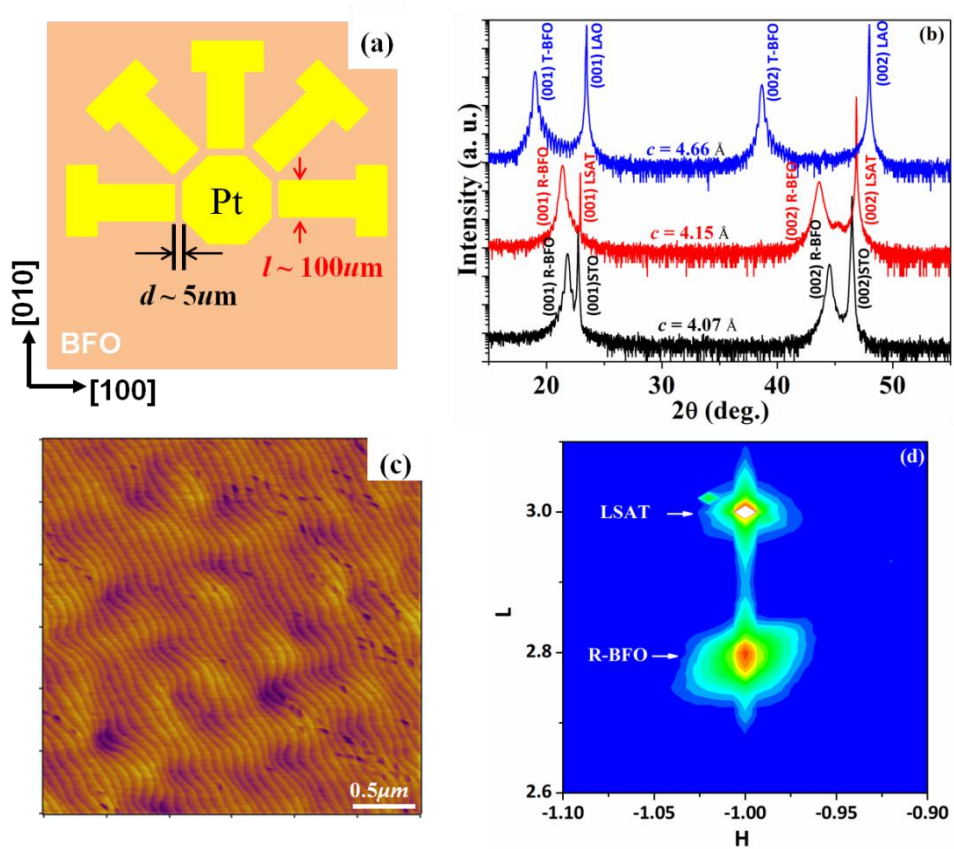

FIG. 1. (a) A schematic of device structure for in-plane *P-E* measurement. (b) XRD θ-2θ scans of ~30-nm-thick BFO films grown on different substrates. (c) AFM topographic image of BFO film grown on LAO. (d) Reciprocal space map around the $\bar{1}03$ reflection of BFO film grown on LSAT.

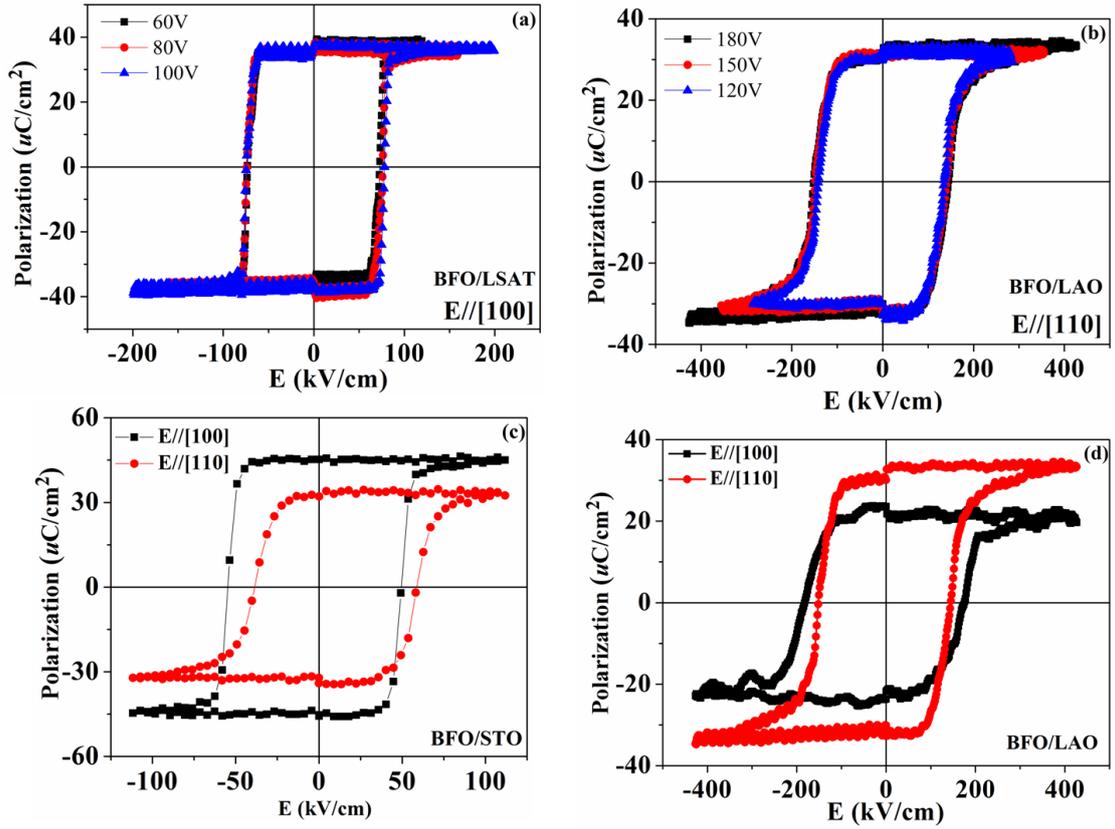

FIG. 2. (a) Field-dependent variation of in-plane *P-E* loops of the BFO film grown on LSAT when the electric field is applied along the [100] direction. (b) Field-dependent variation of *P-E* loops of the pure T-like BFO film grown on LAO, when the applied electric field is along the [110] direction. *P-E* loops of the BFO film grown on (c) STO and (d) LAO along different applied field directions.

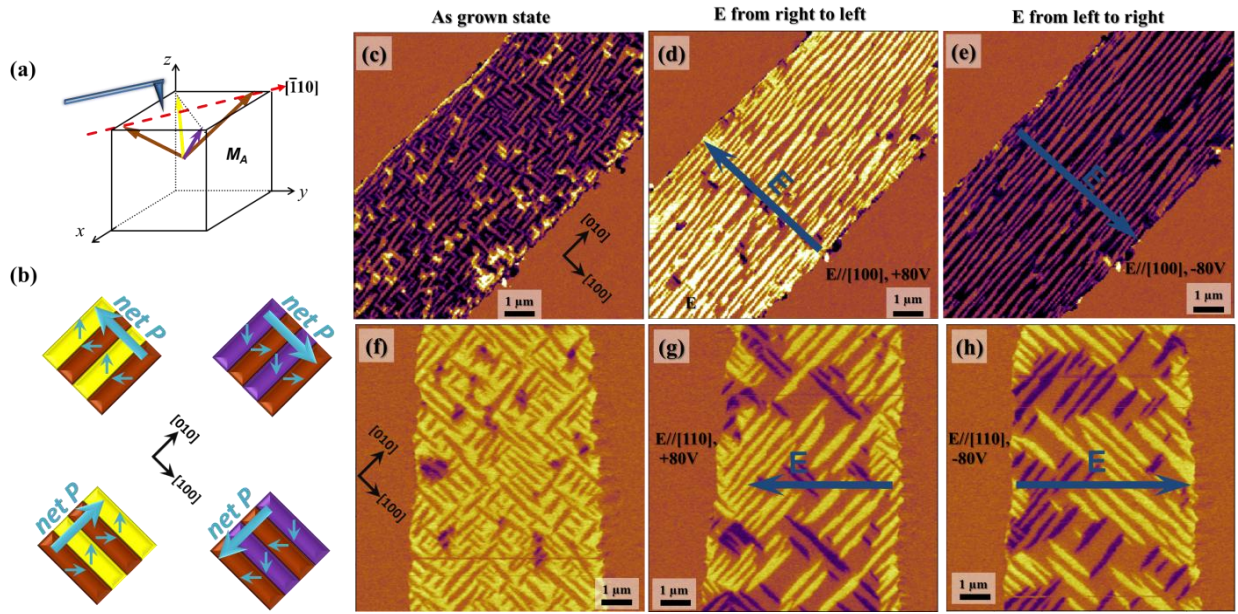

FIG. 3. (a) Illustrations of the unit cell of the R-like $M_A$ phase, and (b): Schematics of in-plane domain configurations of the R-like $M_A$ films. In plane PFM images of quasi-planar BFO capacitors fabricated on films grown on STO substrates with electrode edges along [100] and [110] directions: (c), (f), As-grown state; (d), (g) after electric poling with +80V; and (e), (h) after electric poling with -80V.

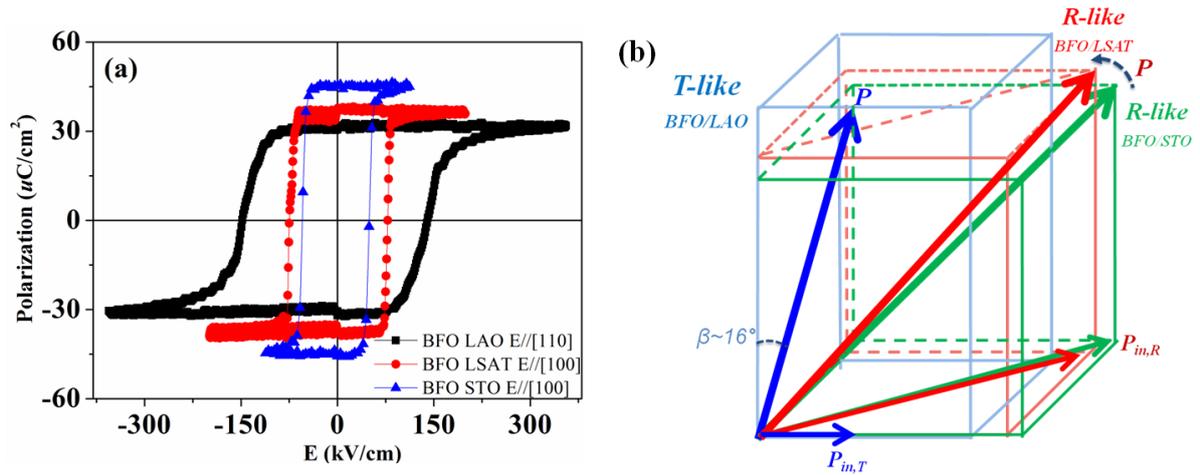

FIG. 4. (a) Comparison of P-E loops of the strained BFO films on different substrates. (b) Schematic of strain-induced spontaneous polarization rotation in the (110) plane of R-like phase and direction of the polarization $P$ in the T-like phase of BFO. The arrows at the end of which the letter $P$ is indicated represent the spontaneous polarization. $P_{in,R}$ and $P_{in,T}$ represent the in-plane component of the polarization in R-like and T-like phase, respectively. $β$ is the angle between the polarization vector of the T-like phase and the [001] direction.

Table I Cartesian Components of the polarization of BFO under different epitaxial strains from first-principles calculations.

| Strain | Phase | Polarization (C/m$^2$) |
|:---:|:---:|:---:|
| 0 | *R3c* | (0.521  0.521  0.520) |
| -1.5% | $M_A$ | (0.470  0.470  0.604) |
| -2.5% | $M_A$ | (0.437  0.437  0.679) |
| -4.5% | $M_A$ | (0.288  0.289  1.386) |
| -4.4% | $M_C$ | (0.000  0.434  1.399) |